\documentclass[twocolumn,showpacs,preprintnumbers,amsmath,amssymb]{revtex4}

\usepackage{graphicx}
\usepackage{dcolumn}
\usepackage{bm}

\begin{document}

\preprint{APS/123-QED}

\title{Self-stabilised fractality of sea-coasts through damped erosion}

\author{B. Sapoval$^{1,2}$, A. Baldassarri$^{1,3}$, A. Gabrielli$^4$}

\affiliation{$^1$Laboratoire de Physique de la Mati\`ere Condens\'ee, C.N.R.S. Ecole Polytechnique, 91128 Palaiseau, France.
\\ $^2$Centre de Mathématiques et de leurs Applications, Ecole Normale Sup\'erieure, 94235 Cachan, France.
\\$^3$INFM, UdR Roma 1, Dipartimento di Fisica, Universit\`a di Roma "La Sapienza", P.le Aldo Moro 2, 00185 Rome, Italy.
\\ $^4$"Enrico Fermi" Center Via Panisperna 89 A, Compendio del Viminale, Palaz. F, 00184 Rome, Italy.}

\date{\today}

\begin{abstract}
Erosion of rocky coasts spontaneously creates irregular seashores.
But the geometrical irregularity, in turn, damps the sea-waves,
decreasing the average wave amplitude. There may then exist a mutual
self-stabilisation of the waves amplitude together with the irregular
morphology of the coast. A simple model of such stabilisation
is studied. It leads, through a complex dynamics of the earth-sea
interface, to the appearance of a stationary fractal seacoast with
dimension close to $4/3$. Fractal geometry plays here the role of a
morphological attractor directly related to percolation
geometry.
\end{abstract}

\maketitle                              

Coastline morphology is of current interest in geophysical research
\cite{eric}, and coastline erosion has important economic consequences
\cite{penning}. Also, the recent concern about global warming has
increased the demand for a better understanding of coastal
evolution. At the same time, although the geometry of seacoasts is
often used as an introductory archetype of fractal morphology in
nature \cite{mandelbrot67,barton}, there has been no explanation about
which physical mechanism could justify that empirical observation. In the 
field litterature\cite{coastalevolution} one can read:
{\em "As a matter of some urgency, researchers concerned with coastal
evolution should consider the alternative models, even if there
are few supporting data. The ideas of ... stochastic development, ... and criticality,
all deserve investigation."} The present work propose a minimal, but
robust, model of evolution of rocky coasts towards fractality.

\begin{figure}
\includegraphics[width=8.0cm,height=8.0cm]{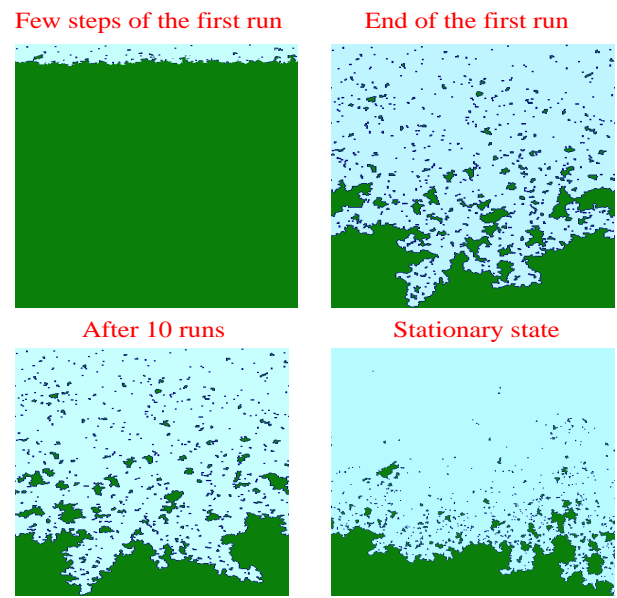}
\caption{\label{fig1}
Time evolution of the coastline geometry. Upper left: situation at the
beginning of the erosion process. The coast is irregular but not yet
fractal. Upper right: the seashore at the end of the ``rapid'' 
mechanical erosion. Lower left and right: situations after 10 and 20
runs of the ``slow'' weathering process (with $\epsilon=0.01$).}
\vspace{-0.5cm}
\end{figure}

The model describes how a stationary fractal
geometry can arises spontaneously from the mutual self-stabilization of
coast morphology and sea eroding power \cite{sapo1}. If, on one hand,
erosion generally increases the geometrical irregularity of the coast,
on the other hand this increase creates a stronger damping of the sea
and a consequent diminution of its eroding power.  The increased
damping argument relies on the studies of fractal acoustical cavities,
which have shown that viscous damping is augmented on a longer,
irregular, surface \cite{sapo2,sapo3}. In the following, a minimal
two-dimensional model of erosion is introduced which leads to the
spontaneous evolution of a smooth seashore towards fractality as shown
in Fig.~\ref{fig1}.

Rocky coasts erosion
is the product of marine and atmospheric causes
\cite{davis}. There exist many different erosion processes: wave quarrying,
abrasion, wetting and drying, frost shattering, thermal expansion,
salt water corrosion, carbonation, hydrolysis. A simplified picture is
used here by assuming that the different processes can be separated
into two categories: ``rapid'' mechanical erosion (namely wave
quarrying) and ``slow'' chemical weakening. The justification is that
mechanical erosion generally occurs rapidly, mainly during storms,
after rock has been altered and weakened by the slow weathering
processes. We first study the supposedly rapid erosion mechanisms. Then we show
that the full complex dynamics, involving fast and slow processes,
build a dynamic equilibrium that preserves the fractal shape of the
coast.

The sea, together with the coast, is considered to constitute a
resonator. It is assumed that there exists an average excitation power of
the waves $P_0$. The ``force'' acting on the unitary length of the
coast is measured by the square of the wave amplitude $\Psi^2$.  This
wave amplitude is related to $P_0$ by a relation of the
type $\Psi^2\sim P_0 Q$ where $Q$ is the morphology dependent 
quality factor of the system;
the smaller the quality factor, the stronger the damping of the
sea-waves. There are several causes to waves damping. Since the
different loss mechanisms occur independently, the quality factor
satisfies a relation of the type
\begin{equation}
\frac{1}{Q}=\frac{1}{Q_{coast}}+\frac{1}{Q_{other}}\,,
\label{eq1}
\end{equation}
where $Q_{coast}$ is the quality factor due to the viscous dissipation
of the fluid moving along the coast and the nearby islands and $Q_{other}$
is related to other damping mechanisms (e.g. bulk viscous damping).
Studies of fractal acoustical cavities \cite{sapo2,sapo3}
 have shown that the viscous damping increases roughly
proportionally to the cavity perimeter. Therefore, one can, in first approximation,
 assume that $Q_{coast}$
is inversely proportional to the coast perimeter $L_p(t)$
whereas $Q_{other}$ is independent of the coast morphology. In other words,
the sea exerts a homogeneous erosion force $f(t)$ on each
coast element proportional to $\Psi^2(t)$:
\begin{equation}
f(t)=\frac{f_0}{1+\frac{g\,L_p(t)}{L_0}}\,,
\label{eq2}
\end{equation}
where $L_p(t)$ is the total length of the coast at time $t$ (then 
$L_p(t=0) =L_0$). The factor $g$ measures the relative contribution to damping of a
flat shore as compared with the total damping. Since $g$ represents the
importance of coast dissipation with respect to the other mechanisms,
which are thought to be dominant, computations are performed under the
condition $g \ll 1$. The quantity $f_0$ is the renormalized value of $P_0$ such
that $f(t)<1$ at all $t$.

\begin{figure}
\includegraphics[width=4.0cm,height=8.5cm,angle=-90]{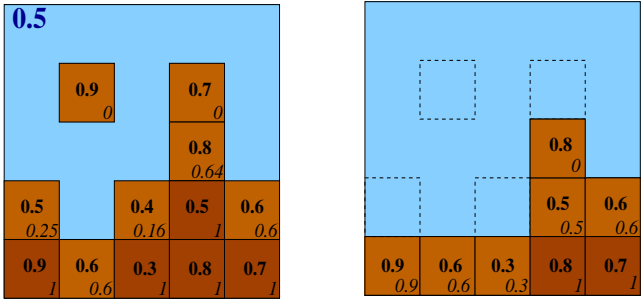}
\caption{\label{fig2}
Illustration of the erosion process. 
The thick numbers at the square centres represent the lithology $\{x_i\}$. 
The numbers in the corners are the corresponding resistances $r_i$ which depend on the 
local environment as explained in the text. The sites marked with 1 in the lower corner are 
earth sites with no contact with the sea. 
On the left and on the right respectively  the situations before and after a single
erosion step with $f(t)=0.5$. After this step resistances are updated due to the new
sea environment.}
\vspace{0.6cm}
\end{figure}

The mechano-chemical properties of the rocks constituting the coast,
which are linked to structure and composition defining their
``lithology", are unknown and exhibit some dispersion. The
``resisting'' random earth is then modelled by a square lattice of
random units of width $L_0$. Each site represents a small portion
of the earth. The sea acts on a shoreline constituted of cells, each
one characterised by a random number $x_i$, between $0$ and $1$,
representing its lithology. The erosion model should also take into
account the fact that a site surrounded by the sea
is relatively weakened as compared with a
site surrounded by earth or other coast sites. Hence, the resistance
to erosion $r_i$ of a site depends on both its lithology and the number
of sides exposed to the action of the sea. This is implemented through
the following weakening rule: sites surrounded by three earth sites
has a resistance $r_i = x_i$. If in contact with $2$ sea sites
the resistance is $r_i = x_i^2$. And, if site $i$ is attacked by $3$
or $4$ sides, it has zero resistance. The iterative evolution rule is simple:
at time $t$, all coast sites
with $r_i < f(t)$ are eroded, and then $L_p(t)$ and $f(t)$ are updated
together with the resistances of the earth sites in contact with the
sea. Then, from one step to the next,
some sites are eroded because they
present a ``weak lithology'' while some strong sites are eroded due to
their weaker stability due to sea neighbouring.  
An example of local evolution is shown in Fig. \ref{fig2}.

\begin{figure}
\vspace{-0.1cm}
\includegraphics[width=7.5cm,height=5.5cm]{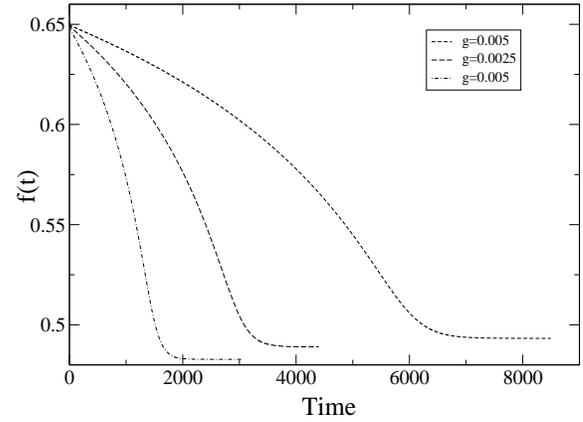}
\caption{\label{fig3}
Time dependence of the erosion force f(t). The figure shows the evolution of the sea 
erosion force acting on the coastline, for $f(t=0) =0.65$ and different values of the 
scale gradients g. The dynamics spontaneously stops at a value weakly dependent on g. 
Data refer to 
systems with $L_0 = 1000$. Each curve is obtained by averaging 
over ten different realisations of the earth lithology distribution.}
\vspace{-0.5cm}
\end{figure}

\begin{figure}
\includegraphics[width=6.0cm,height=8.5cm,angle=-90]{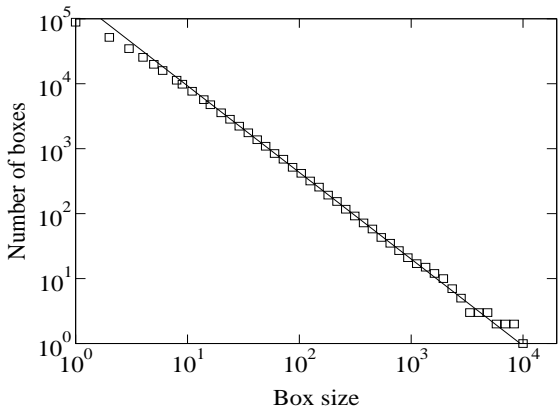}
\caption{\label{fig4}
Box-counting determination of the coast fractal dimension. The straight line is a 
power law with a slope -4/3. The data refer to a system with $L_0 = 10^4$, and 
$g =10^{-4}$.}
\end{figure}

In the first steps of the dynamics, the erosion front keeps quite smooth and it roughens
progressively as shown in Fig. \ref{fig1} upper row. During the process,
finite clusters are detached from the infinite earth, creating
{\em islands}. At any time, both the
islands and the coastline perimeters contribute to the 
damping. As the total coastline length $L_p(t)$ increases, the sea force
becomes weaker. At a certain time step $t_f$, the weakest point of the
coast is stronger than $f(t_f)$ and the dynamics stops. This shows
that erosion reinforces the coast by preferential elimination of its
weakest elements until the coast is strong enough to resist further
erosion. The time evolution of $f(t)$ is shown in Fig. \ref{fig3}. 
The fact that the process spontaneously leads to a total stop
of erosion for a \emph{non-zero} wave power is remarkable. 
It is a direct consequence of random percolation
\cite{stauffer} as discussed below.
At time $f(t_f)$ the coastline is
fractal (see Figs. \ref{fig1} and \ref{fig4} ) with dimension $D_f = 1.332(3)$, very
close to 4/3, up to a characteristic width $\sigma$. This width
$\sigma$ is defined as the standard deviation of the final coastline
depth $\left<(y - Y)^2\right>^{1/2}$ (where $Y = \left<y\right>$ is the
average erosion depth). Fig. \ref{fig4} shows the box-counting determination of
the fractal dimension. 
The fact that $D_f \simeq 4/3$
confirms the relation with percolation, as
$4/3$ is the dimension of the
so-called accessible percolation cluster \cite{grossman}. Note that this 
value is also close to the observed fractal dimension for many seacoasts, 
as the United States eastern shore \cite{barton}. 

\begin{figure}
\includegraphics[width=6.0cm,height=8.5cm,angle=-90]{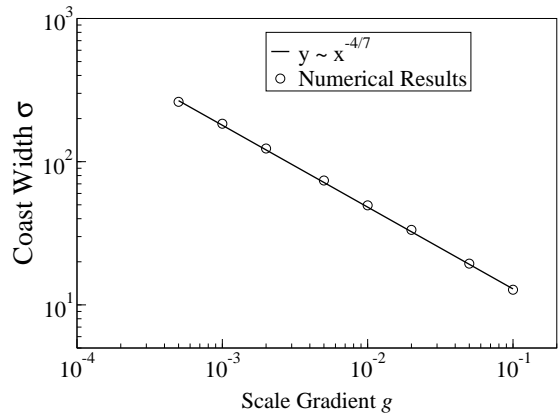}
\caption{\label{fig5} Scaling behaviour of the coast width
$\sigma$. The straight-line is a power law fit with the GP exponent
-4/7. Each point is an average over $400$ samples 
with $L_0 = 5000$,
$f(0)=0.55$.}
\end{figure}

A more detailed study
of the coast width $\sigma$ suggests that the model falls in
the universality class of Gradient Percolation (GP) \cite{sapo4}. As shown in
Fig. \ref{fig5}, the coast width $\sigma$ follows a scaling law with respect
to $g$
\begin{equation}
\sigma\sim g^{-\alpha_\sigma}
\label{eq3}
\end{equation}
with $\alpha_\sigma \simeq 4/7$. 
This law is characteristic of GP where
sites are occupied at random with an occupation probability that
varies in space along one fixed direction 
between 1 and 0 with a constant gradient $g$, and $\sigma$ is
the width of the frontier of the infinite cluster. 
The fact that here $g$ is proportional to a gradient of occupation
probability is explained by the following argument. At time $t$,
the erosion power is $f(t)$ while the sea has eroded the earth up to an
average depth $Y(t)$, an increasing function of $t$. Inverting this function, $f$ 
can be written as
$f(t(Y))$. There exists then a spatial gradient of the
occupation probability by the sea. For small enough $g$ one
can write $|df/dY | = |(g/L_0) (dL_p(Y)/dY)|$. The quantity $dL_p(Y)/dY$ is
a function of $g$ but to the lowest order it is a constant independent
of $g$ since even with $g = 0$, there will be an erosion due to randomness
and a consequent perimeter evolution $L_p(t)$. Then to lowest order, the
real gradient $df/dY$ is linear in $g$, which can then be called the 
``scale gradient''.  Note that the formation of a fractal interface
due to the spontaneous appearance of a gradient has already been
observed in the corrosion of an aluminium film \cite{balasz,sapo5}.

Of course, the real dynamics of the coasts are more complex than the ``rapid''
process considered above. They result from the
interplay with the slow weathering processes, generally attributed to
carbonation or hydrolysis. These processes act on longer, geological,
time scales. In order to simulate this evolution, the lithology parameter $x_i$ 
of all the coast sites is decreased by a small fraction $\epsilon$,
i.e. $x_i'=(1-\epsilon)x_i$ with $\epsilon\ll 1$ after the erosion has stopped at 
$t_f$. One or a few coast
sites then become weaker than $f(t_f)$ and the rapid erosion
dynamics starts again up to a next arrest time. This procedure
is then iterated. At each restart of erosion, a finite portion of the
earth is eroded. The repeated effect of the ``slow'' weathering
mechanisms gives a fluctuating behavior of $f(t)$ and
generates a coastline erosion drift with fluctuations or
avalanches, but it does not alter the fractal features of the
coast (see Figure 1 lower row). In that sense fractal geometry plays
the role of a statistical attractor.  In the language of coastal
studies \cite{haslett}, the system state evolves through a dynamical equilibrium
where small perturbations (even without special trigger events) may
stimulate large fluctuations and avalanche dynamics. This is due to
the underlying criticality of percolation systems. 
In the authors' mind, it is this stationary regime which
corresponds to the geomorphologic observations of fractal
seacoasts.

The above simple model presents limitations, but the simplifications
that keep the system in the universality class of percolation are
unimportant. The use of different weakening rules would eventually
modify the dynamics but not the final fractal dimension. The time
separation between rapid erosion and slow chemical weakening is
somewhat arbitrary as both mechanisms can occur simultaneously without
changing the coast fractality. Also a better model for damping should
take care of a wave frequency dependence as well as it should consider
the existence of localization by the frontier of the waves along the
irregular coast \cite{sapo2}. This would modify Eq. (2) and change the
time evolution. However, since percolation possess the universality
properties of phase transitions \cite{stauffer}, the fractal dimension
of the coast should not depend on these factors. In the context of
corrosion dynamics \cite{gab1}, this has been shown through arguments
from dynamical field theory of absorbing states
\cite{gab2}. 
If large-scale modifications or correlations exist in the lithology
properties, then the resulting geometry would be more complex. But our
main result, namely the existence of irregular coasts as a result of a
self-stabilisation mechanism, would remain correct even though the
geometry would be more complex than that of critical percolation. We
believe that it is in those terms that can be interpreted the results
of several detailed study of the self-similarity properties of
sea-coasts \cite{goodchild,andrle}.

A more radical change in the geometry is expected if sediments
transport is taken into account. In our simplified model, the
sediments are supposed to be transported offshore and disappear, while
in the real erosion process these sediments are partially transported
along the coast. However our approximation is justified by some
examples, like the North Coast of Oregon (USA), a leading hedge coast
dominated by erosion with very limited coastal deposits
\cite{haslett}.

One should also mention that it has been found very recently that the
GP power laws apply even when the front is too narrow to be fractal
\cite{sapo6}. This extends considerably the range of application of GP
to rocky coasts which are irregular, but not to a fractal range. 

This work has presented a minimal model for the formation of fractal rocky
coast morphology. This model bears on the reciprocal evolution of the
erosion power and the topography of the coast submitted to that
erosion: The more irregularly eroded the coast is, the weaker the
average sea erosion power. Note that this seems to be an
empirically known effect used to build efficient break-waters that are
based on hierarchical accumulations of tetrapods piled over layers of
smaller and smaller rocks, in close analogy with fractal geometry
\cite{manual}. 
The retroaction leads to the spontaneous formation of a fractal
seacoast with a fractal dimension $D_f = 4/3$. The fractal geometry
plays here the role of a morphological attractor: whatever its initial
shape, a rocky shore will end fractal when submitted to such type of
erosion, forgetting its initial morphology.  Note that our model
suggests that, on the field, the islands which have resisted to an
erosion under a force larger that the final force $f(t_f)$, are
stronger that the coast itself. This could be verified on the
historical data of known seacoasts and neighbouring islands
evolutions.  The model reproduces at least qualitatively some of the
features of real coasts using only simple ingredients: the randomness
of the lithology and the decrease of the erosion power of the
sea. It is worth to be noted that the use of
simple geophysical ingredients leads to an evolution towards a
self-organized fractality directly related to percolation theory.

A.G. wish to acknowledge the Departement of Physics of
the University of Rome ``La Sapienza'' (Italy) for supporting this research.

\end{document}